\documentclass[reviewcopy]{elsarticle}

\usepackage[reviewcopy]{adndt}
\usepackage{longtable,color}
\usepackage{booktabs}

\usepackage{amsmath}

\usepackage{amssymb}

\usepackage{color}
\usepackage[usenames,dvipsnames]{xcolor}

\biboptions{square,sort&compress}
\bibpunct[]{[}{]}{,}{n}{}{;}
\begin{document}

\begin{frontmatter}

\journal{Atomic Data and Nuclear Data Tables}

%% Author, fill in article title here

\title{Isotope Shifts in Beryllium-, Boron-, Carbon-, and Nitrogen-like Ions from Relativistic Configuration Interaction Calculations}

%% Fill in author list here
\author[ulb]{C. Naz\'e}
\author[ulb]{S. Verdebout}
\author[lithuania2]{P. Rynkun}
\author[lithuania2]{G. Gaigalas} 
\author[ulb]{M. Godefroid}\corref{cor1}
  \ead{mrgodef@ulb.ac.be}
%  \ead{E-mail: user@some.where.net}
\author[sweden]{P. J\"onsson}

\cortext[cor1]{Corresponding author.}
%  \fntext[X]{First author footnote.}
%  \fntext[Y]{Second author footnote.}

  \address[ulb]{Service de Chimie Quantique et Photophysique, CP160/09, \\ 
Universit\'e Libre de Bruxelles, 
Avenue F.D. Roosevelt 50, B~1050 Brussels, Belgium}
\address[lithuania2]{Vilnius University, Institute of Theoretical Physics and Astronomy, LT-01108 Vilnius, Lithuania}
\address[sweden]{Group for Materials Science and Applied Mathematics, Malm\"{o} University, 205-06 Malm\"{o}, Sweden}

\date{16.12.2002} %please do not use \today, use actual date of version

\begin{abstract}  
Energy levels, normal and specific mass shift parameters as well as electronic densities at the nucleus are reported for numerous states along the beryllium, boron, carbon, and nitrogen isoelectronic sequences. Combined with nuclear data, these electronic parameters can be used to determine values of level and transition isotope shifts. The calculation of the electronic parameters is done using  first-order perturbation theory with relativistic configuration interaction wave functions that account for valence, core-valence and core-core correlation effects as zero-order functions. Results are compared with experimental and other theoretical values, when available.
\end{abstract}

\end{frontmatter}

%%% Keywords and subject classification are not used in ADNDT 
%%%\begin{keywords}
%%%Insert list of keywords here.
%%%\end{keywords}

%%% The table of contents should start a new page. This command will
%%% automatically pull all the unstarred \section, \subsection and
%%% \subsubsection titles into the Contents. Starred versions need to be
%%% done manually using
%%%            \addcontentsline{toc}{[[sub]sub]section}{Section title}
%%% at the correct place. Examples are given below.

%%% The lists of data figures and data tables are created automatically
%%% by the \listofDfigures and \listofDtables commands.

\newpage

\tableofcontents
\listofDtables
\listofDfigures
\vskip5pc
\newpage

\section{Introduction}

High resolution solar and stellar spectra reveal isotope shifts (IS) and hyperfine structures of many spectral lines. These small structures are not always completely resolved and instead they shift and
broaden the atomic lines. To correctly interpret the spectra it is often necessary to
include isotope shifts and hyperfine structures in a theoretical modeling of the line profiles~\cite{Kur:93a,Cayetal:2007a}. Hyperfine and isotope data are also needed for generating synthetic spectra to study isotope anomalies
in astronomical objects~\cite{Proetal:99a} and for understanding nucleosynthesis mechanisms~\cite{Lunetal:2007b,Roeetal:2012a}.
In the last decade, several studies have shown that the typical isotope shifts may be of the same order of magnitude as the effects of the possible temporal and spatial variation of the fine-structure constant. In this framework, Kozlov \textit{et al.}~\cite{Kozetal:2004a} proposed a method to exclude systematic effects caused by changes in the isotope abundances during the evolution of the universe, but there is a need 
to develop methods that can accurately estimate isotope shifts~\cite{Beretal:08a}.
Isotope shifts of spectral lines are also used in nuclear physics to estimate the
root-mean-square nuclear charge radii and mean-square  radii changes~$\delta\langle r^2\rangle^{A,A'}$ for which the electronic factors play a crucial role~(see for example~\cite{Braetal:2008a,Noretal:2011a,Canetal:2012a,Pucetal:2013a,AngMar:2013a}).

Relativistic effects on the electronic structure are expected to increase with the nuclear charge. Shabaev~\cite{Sha:85a,Sha:88a} 
and,  in an independent way, Palmer~\cite{Pal:87a}, derived relativistic corrections to the mass shift.
Since then several papers have shown the importance of relativistic effects on  isotope shifts~\cite{Tupetal:2003a,KorKoz:2007a,Kozetal:2008a,Poretal:2009a, Kozetal:2010a}. The development of experimental techniques allowing the accurate measurement of line IS enhances the need for accurate calculations of the relevant electronic parameters that account for both electron correlation and relativistic effects, especially for medium and heavy open-shell atoms \cite{Proetal:2012a, Cheetal:2012a}. 
Amongst these techniques, the storage ring measurement of isotope shifts in the spectrum of resonant electron-ion recombination of heavy few-electron ions gives access to nuclear charge radii changes involving heavy stable and unstable nuclides, as demonstrated for three-electron Nd$^{57+}$ isotopes produced at GSI's storage ring ESR~\cite{Braetal:2008a}. Dielectronic recombination (DR) measurements have also been performed at the heavy-ion storage ring TSR  in Heidelberg for Be-, B- and C-like ions relevant to astrophysics and plasma physics~\cite{Sch:2009a,Sch:2012a}, although no IS in the DR spectra have been reported so far. DR experiments are also envisaged at ISOLDE \cite{Grietal:2012a} where the availability of radioisotopes together with the TSR electron collision facilities will open up new opportunities. 

\textsc{Grasp2K} \cite{Jonetal:2007a} is a fully relativistic code based on the multiconfiguration Dirac-Hartree-Fock (MCDHF) method. Its latest release \cite{Jonetal:2013a} includes a module, \textsc{ris3}~\cite{Nazetal:2013a}, that evaluates the isotope shift parameters including the relativistic corrections to the recoil operator as derived by Shabaev~\cite{Sha:85a,Sha:88a}.
The purpose of the present work is 
to complement the available atomic data sets along the beryllium, boron, carbon and nitrogen isoelectronic 
sequences with accurate IS electronic parameters that can be used for modeling high resolution stellar spectra, and for interpreting future measurements of isotope effects that would require the knowledge of the electronic factors to access the nuclear charge radii.
\section{Isotope shift theory}
An atomic spectral line is characteristic of the element producing the spectrum. 
The frequency of a spectral line $k$ connecting levels $\ell \leftrightarrow u$, with $E_u > E_\ell$, is given by
\begin{equation}
\nu_k  = \frac{E_u - E_\ell}{h} \; ,
\end{equation}
and differs from one isotope to another. The line frequency isotope shift between isotopes $A$ and $A'$ (with $A > A'$) is
\begin{equation} \label{eq:line_freq_IS}
\delta \nu_{k} ^{A,A'} \equiv \nu_{k}^A - \nu_{k}^{A'} = \frac{\delta E_u^{A,A'}  - \delta E_\ell^{A,A'} }{h} \; ,
\end{equation}
with
\begin{equation}
\delta E_i^{A,A'}  = E_i^A - E_i^{A'}  \hspace*{0.5cm} ( i=\ell, u) \; .
\end{equation}
The isotope shift  is made up of two effects: the mass shift (MS) and the field shift (FS). Generally, the FS dominates for heavy atoms
whereas the MS plays a prominent role for light atoms. In the present work, both effects are expressed as expectation values of operators that are calculated through first-order perturbation theory~\cite{Bluetal:87a} using approximate solutions of the fully relativistic Dirac-Hartree-Fock(-Breit) Hamiltonian as zero-order wave functions.
 
\subsection{Mass shift}
The finite mass of the nucleus gives rise to a recoil effect, called the mass shift.
The nuclear recoil corrections within the $(\alpha Z)^4m^2/M$ approximation are obtained by evaluating the expectation values of the operator \cite{ShaArt:94a,Tupetal:2003a}
\begin{eqnarray}
\label{eq:H_MS}
   {H}_{\rm MS}  \; = \;
 \frac{1}{2M}
 \sum_{i,j} ^{N} \;
\left( {\bf  p}_{i} \cdot {\bf  p}_{j} - \frac{\alpha Z}{r_i} \mbox{\boldmath $ \alpha$}_{i} \cdot  {\bf  p}_j- \frac{\alpha Z}{r_i} \frac{\left(\mbox{\boldmath $ \alpha$}_i\cdot{\bf  r}_i\right) {\bf  r}_i}{r^{2}_{i}}\cdot  {\bf  p}_j\right)\;.
\end{eqnarray}
Separating the one-body ($i=j$) and two-body terms ($i \neq j$) into the normal mass shift (NMS)  and specific mass shift (SMS) respectively, the expression~(\ref{eq:H_MS}) splits in  
\begin{eqnarray}
\label{eq:separation}
{H}_{\rm MS} = {H}_{\rm NMS} + {H}_{\rm SMS} \; .
\end{eqnarray}
For a given atomic state $i$, the (mass-independent) normal mass shift $K_{\rm NMS}$ and specific mass shift $K_{\rm SMS}$ parameters are defined by the following expressions
\begin{eqnarray}
\label{eq:NMS}
  K_{i, \rm NMS}\; &\equiv& M   \langle \Psi_i|
   {H}_{\rm NMS}| \Psi_i \rangle \;,     \\
\label{eq:SMS}
   K_{i, \rm SMS}  &\equiv & M
   \langle \Psi_i |
  {H}_{ \rm SMS}| \Psi_i \rangle \;.%\\    
\end{eqnarray}
These mean values can be expressed as
\[
K_{\rm NMS}=K^1_{\rm NMS} + K^2_{\rm NMS} + K^3_{\rm NMS} \]
 and 
 \[
 K_{\rm SMS}=K^1_{\rm SMS} + K^2_{\rm SMS} + K^3_{\rm SMS} \; ,
 \] where each contribution refers to one of the three terms appearing in (\ref{eq:H_MS}). 
More complete developments, as well as the derivation of the tensorial form of (\ref{eq:H_MS}), are detailed in the work of Gaidamaukas \textit{et al.}~\cite{Gaietal:2011a}. 

 The corresponding level mass shift between two isotopes $A$ and $A'$ with masses $M$ and $M'$, respectively, can be written
\begin{eqnarray}
\delta E^{A,A'}_{i, \rm MS}=\left(\frac{1}{M}-\frac{1}{M'} \right) \left(K_{i, \rm NMS}+K_{i, \rm SMS}\right)\;, \; \; \; \text{with}\; \; \; A > A'.
\end{eqnarray}
In atomic units, the nuclear masses $M$ and $M'$ that are usually given in units of the unified atomic mass (u), must be converted to atomic units of mass ($m_e$) to be consistent with $K$ factors expressed in $[m_e E_{\rm h}]$ units.
For discussing the transition isotope shift  (\ref{eq:line_freq_IS}), one needs to consider the variation of the mass parameter from one level to another.
The corresponding line  frequency isotope mass shift  is  written as the sum of the NMS and SMS contributions:
\begin{eqnarray} \label{split_freq_mass}
\delta \nu_{k,{\rm MS}}^{A,A'}  = \delta \nu_{k,{\rm NMS}}^{A,A'}+\delta \nu_{k,{\rm SMS}}^{A,A'} \;,
\end{eqnarray}
with 
\begin{eqnarray}\label{Delta_nu_MS}
\delta \nu^{A,A'}_{k,{\rm MS}}&=&\left(\frac{M'-M}{MM'}\right)\frac{\Delta K_{\rm MS}}{h}
=\left(\frac{M'-M}{MM'}\right)\Delta \widetilde K_{\rm MS},
\end{eqnarray}
where  $\Delta K_{\rm MS} =  (K_{u,\rm MS} -  K_{\ell,\rm MS} )$ is the difference of the $K_{\rm MS}$ $(=K_{\rm NMS} + K_{\rm SMS} )$ parameters of the upper~($u$) and lower~($l$) levels involved in the transition~$k$. For $\Delta \widetilde K_{\rm MS}$, the unit (GHz~u) is often used in the literature. As far the conversion factor is concerned, we use  
$\Delta K_{\rm MS}$[$m_e E_{\rm h}]= 3609.4824~\Delta \widetilde K_{\rm MS}[{\rm GHz~u}]$.

\subsection{Field shift}
The energy shift arising from the difference in nuclear charge distributions between two isotopes  $A$ and $A'$ for a given atomic state $i$ is called the level field shift. 
Neglecting higher order isotopic variation of the nuclear charge distribution, the level field shift can be written~\cite{Toretal:85a,Bluetal:87a}
\begin{equation}\label{eq:LSF_fact}
 \delta E_{i,\rm FS} ^{A,A'} =  (h \widetilde {\cal F}_i  ) \; \delta\langle r^2\rangle^{A,A'}, 
\end{equation}
where 
\begin{equation}\label{delta_r_square}
 \delta\langle r^2\rangle^{A,A'}  =   \langle r^2\rangle^{A} -  \langle r^2\rangle^{A'} 
\end{equation}
is the difference of the nuclear root-mean-square (rms) charge radii involved. The level field shift electronic factor $ \widetilde {\cal F}_i $ is given (in units of frequency divided by a length squared) by 
\begin{equation} \label{level_electronic_factor}
\widetilde {\cal F}_i= \frac{2\pi}{3 h}Z  \left( \frac{e^2}{4 \pi \epsilon_0} \right) \;  |\Psi(0)|^2_i  \;, 
\end{equation}
where 
$  |\Psi(0)|^2 $ is the total probability density at the origin that can  be  estimated by taking the (${\bf r} \rightarrow {\bf 0}$) limit of the electron density~\cite{Nazetal:2013a} (see section~3.1)
\begin{equation} \label{el_dens_dif_1}
 |\Psi(0)|^2_{i}   =  \lim_{{\bf r} \rightarrow {\bf 0}} \rho^{e}_{i} ( {\bf r} ) =  \frac{1}{4 \pi}  \lim_{r \rightarrow 0} \rho^{e}_{i} ( {r} ) \; .
\end{equation}
 Using~(\ref{eq:LSF_fact}) for levels $i= (\ell, u)$ involved in transition $k$, 
the frequency field shift of the spectral line $k$ can be written as~\cite{Frietal:95a,Toretal:85a,Bluetal:87a}
\begin{equation} \label{Line_FS}
 \delta \nu_{k,{\rm FS}}^{A,A'} =   \frac{\delta E_{u,{\rm FS}} ^{A,A'} - \delta E_{\ell,{\rm FS}} ^{A,A'} }{h} =  F_k  \; \delta\langle r^2\rangle^{A,A'} \; .
\end{equation}
$F_k$ is the line electronic factor (in units of frequency divided by a length squared)
\begin{equation} \label{electronic_factor}
F_k  =  \frac{2\pi}{3h} Z \left( \frac{e^2}{4 \pi \epsilon_0} \right)  \Delta |\Psi(0)|^2_k  \; ,
\end{equation}
proportional to the change  of the total probability density at the origin
associated with the electronic transition between levels $\ell$ and $u$. 
In this approximation, the first-order level- and  frequency-field shifts (\ref{eq:LSF_fact}) and (\ref{Line_FS}) 
 become
\begin{equation}\label{eq:LSF_fact_sm_neg}
\delta E_{i,{\rm FS}} ^{A,A'}  =(h \widetilde {\cal F}_i  ) \; \delta\langle r^2\rangle^{A,A'}= \frac{2\pi}{3}Z  \left( \frac{e^2}{4 \pi \epsilon_0} \right) \;  |\Psi (0)|^2_i \; \delta\langle r^2\rangle^{A,A'} \;,
\end{equation}
and
\begin{equation}\label{Line_FS_sm_neg}
\delta \nu^{A,A'}_{k,{\rm FS}} = F_k  \; \delta\langle r^2\rangle^{A,A'}= \frac{Z}{3 \hbar} \left( \frac{e^2}{4 \pi \epsilon_0} \right) \;  \Delta |\Psi (0)|^2_k  \; \delta\langle r^2\rangle^{A,A'} \;,
\end{equation}
respectively. The wavenumber ($\sigma= E/hc$) field shift
\begin{equation}\label{Line_FS_sm_cm}
\delta \sigma^{A,A'}_{k,{\rm FS}} =  \frac{Z}{3 \hbar c} \left( \frac{e^2}{4 \pi \epsilon_0} \right) \; \Delta |\Psi (0)|^2_k  \; \delta\langle r^2\rangle^{A,A'} \;,
\end{equation}
is also often used by experimentalists. 

\subsection{Total line frequency shift}
One can estimate the total line frequency  shift by merely  adding the mass shift~(\ref{Delta_nu_MS}) and field shift~(\ref{Line_FS}) 
 contributions\footnote{The line frequency shift~(\ref{split_freq_tot}) is often written as 
$\delta \nu_k^{A,A'}  = M_k \;  \frac{A'-A}{AA'} +  F_k \; \delta \langle r^2 \rangle^{A,A'} $ }
\begin{eqnarray} \label{split_freq_tot}
\delta \nu_k^{A,A'} & = &\underbrace{\delta \nu_{k,{\rm NMS}}^{A,A'}+\delta \nu_{k,{\rm SMS}}^{A,A'}}_{{\delta \nu_{k,{\rm MS}}^{A,A'}}}+\delta \nu_{k,{\rm FS}}^{A,A'} \nonumber \\
 & = & \left(\frac{M'-M}{MM'}\right)\Delta \widetilde K_{\rm MS}  +  F_k  \; \delta\langle r^2\rangle^{A,A'} \; .
\end{eqnarray}

\section{Computational procedure}
\subsection{Multiconfiguration Dirac-Hartree-Fock}
The multiconfiguration Dirac-Hartree-Fock (MCDHF) method~\cite{Gra:2007a}, as implemented in the program package \textsc{Grasp2K}
\cite{Jonetal:2007a,Jonetal:2013a},
is used to obtain approximate wave functions describing the atomic levels.
 An atomic state function (ASF) is represented by a linear combination of configuration state
functions (CSFs) with same parity $P$, total angular momentum $J$, and its component along $z$-direction, $M_J$,
\begin{equation}
\label{ASF}
\Psi(\gamma PJM_J)   = \sum_{j=1}^{NCSF} c_{j} \Phi(\gamma_{j}PJM_J),
\end{equation}
where $\{c_j\}$ are the mixing coefficients and $\{\gamma_j\}$
the sets of configuration  and intermediate shell-coupling quantum numbers needed to unambiguously specify   the CSFs. 
The latter are built on single-electron Dirac spinors. 
The mixing coefficients and the single-electron orbitals (radial large and small components) are obtained by iteratively solving  the relativistic self-consistent field (RSCF) equations and the secular equation associated with the configuration interaction matrix. A more limited version is the relativistic configuration interaction (RCI) approach using a fixed pre-optimized set of orbitals and allowing only the mixing coefficients to be varied. In the RCI computation, the Breit interaction and leading QED effects (the vacuum polarization and self-energy) can be taken into account perturbatively as
well. It should be emphasized that the use of a point-like nucleus gives unreliable electron densities at the nucleus in the relativistic scheme. A finite nuclear charge distribution model should be used instead to ensure the first-order perturbation to be valid for the field shift~\cite{Bluetal:87a}.  In the present work, the
two-parameters $(c,a)$ Fermi nuclear model
\begin{equation}
\rho(r) = \frac{\rho_0}{1 + e^{(r-c)/a}}
\end{equation}
is adopted to generate the nuclear potential felt by the electrons. 
Here $\rho_0$ is a normalization coefficient, $c$ is the half-density radius and $a = t/(4 \ln 3)$ is related to the surface
thickness $t$ of the charge distribution. In practice, the value $t = 2.30$ fm is used and $c$ is computed according to the
formulae given in \cite{ParMoh:92a}. Since the ASFs are relatively insensitive to the details of the nuclear model,
the nuclear parameters from any stable isotope of the considered chemical element $_Z$X can be safely chosen for performing the MCDHF and RCI calculations of the electronic wave functions. As discussed by Blundell {\it et al.}~\cite{Bluetal:87a}, and as confirmed by more recent studies (see for instance \cite{Lietal:2012a}), the error inherent to the use of a first-order perturbation picture that neglects changes in the electron wave functions arising from variations in the nuclear charge distribution, is estimated of the order of one part in  $10^3$-$10^4$.
\subsection{Computation of isotope shift parameters}
The isotope shift parameters and the electron density at the nucleus are calculated in a first-order perturbation approach using the MCDHF or RCI atomic state functions as the zero-order wave functions. Rewriting the operators for the normal mass shift parameter $K_{\rm NMS}$, the specific mass shifts parameter $K_{\rm SMS}$ 
as well as for the electron density in tensorial form~\cite{Gaietal:2011a}, the calculation of these quantities reduces to a summation over 
matrix elements between CSFs. These matrix elements, in turn, are expressed as sums over radial integrals weighted by angular factors.
All isotope shift electronic parameters were calculated using the  isotope shift module \textsc{ris3}~\cite{Nazetal:2013a} interfaced with the \textsc{Grasp2K} package~\cite{Jonetal:2013a}. Angular factors were saved on disk and re-used for all ions in an isoelectronic sequence to reduce the execution time.  
\section{Generation of configurations expansions}

MCDHF and RCI calculations can be performed for single levels, but also for portions of a spectrum.  In this work calculations were carried out by parity and configuration, that is, wave functions for all states belonging to a specific configuration were determined simultaneously in an EOL calculation~\cite{Dyaetal:89a}. Using the latter scheme, a balanced description of a number of fine-structure states belonging to one or more configurations can be obtained in a single calculation. 
The expansions for the ASFs were obtained using the active set method \cite{Sturetal:2007a}. 
Here CSFs of a specified parity and $J$-symmetry are generated by excitations from a number of reference configurations to a set of relativistic orbitals. By applying restrictions on the allowed excitations, different electron correlation effects can be targeted. To monitor the convergence of the calculated energies and the physical quantities of interest, the active sets are increased in a systematic way by progressively adding layers of correlation orbitals. 

\subsection{Boron-, carbon-, and nitrogen-like systems} 
The starting point was a number of MCDHF calculations where the CSF expansions for the states belonging to a configuration were obtained by single and double (SD) excitations from a multireference (MR) consisting of the set of configurations that can be formed by the same principal quantum numbers as the studied configuration,
to an increasing active set of orbitals. The MCDHF calculations were followed by RCI calculations including the Breit
interaction and leading QED. The CSF expansions for the RCI calculations were obtained by SD-excitations from  a larger MR, including the most 
important configurations outside the complex, to the largest active set of orbitals. The correlation models used for the boron-, carbon-, and nitrogen-like ions are summarized in Table~A. The reference configuration is shown on the left. Columns two and three display the MR for, respectively, the MCDHF and RCI expansions. The largest active set is shown in column four, where the number of orbitals for each $l$-angular symmetry is specified.
The last column gives the number of CSFs for the RCI calculations. The described computational strategy has previously been used by Rynkun \textit{et al.}~\cite{Rynetal:2012a,Jonetal:2011a,Rynetal:2013a} to compute energies and transition rates, and more details  can be found in these papers. 

\subsection{Beryllium-like systems}
The beryllium-like systems are a little different. For each parity, the wave functions for all
states belonging to the $1s^22s^2$ and $1s^22p^2$ even configurations, and to the $1s^22s2p$ odd configuration, were determined simultaneously. 
For the MCDHF calculations, the CSF expansions were obtained by merging CSFs generated by SD-excitations from the reference configurations to an increasing active set with CSFs  obtained by single, double, triple, and quadruple (SDTQ) excitations to a subset of the active set of orbitals. The largest active set consisted
of orbitals with principal quantum numbers $n\leq8$ and the subset was formed by orbitals with principal quantum numbers $n\leq 4$. The MCDHF calculations were followed by RCI calculations including the Breit
interaction and leading QED. The expansions for the RCI calculations were obtained by merging CSFs generated by SDTQ-excitations from the reference configurations to the largest active set with the restriction that there in each CSF is at least two orbitals with $n \leq 3$ with CSFs generated by
SDTQ-excitations from the reference configurations to the active set of orbitals with principal quantum numbers $n\leq4$. The RCI expansions included about 296~000 relativistic CSFs. 

\begin{table}
\caption{Generated CSF expansions for the MCDHF and RCI calculations for the boron-, carbon-, and nitrogen-like ions.}
\begin{tabular}{llllr} \hline
Configuration         & MR for MCDHF             & \multicolumn{1}{l}{MR for RCI}                  & \multicolumn{1}{c}{Active set}  & \multicolumn{1}{c}{NCSF in RCI} \\ \hline
\multicolumn{5}{c}{boron-like} \\ \hline
$1s^22s^22p$   & $1s^2\{2s^22p,2p^3\}$   & $1s^2\{2s^22p,2p^3,2s2p3d,2p3d^2\}$             & $\{9s8p7d6f5g3h1i\}$ &   200~10$^3$ \\
$1s^22p^3$     & $1s^2\{2s^22p,2p^3\}$   & $1s^2\{2s^22p,2p^3,2s2p3d,2p3d^2\}$             & $\{9s8p7d6f5g3h1i\}$  &   360~10$^3$ \\
$1s^22s2p^2$     & $1s^22s2p^2$            & $1s^2\{2s2p^2,2p^23d,2s^23d,2s3d^2\}$           & $\{9s8p7d6f5g3h1i\}$  &   300~10$^3$ \\ \hline
\multicolumn{5}{c}{carbon-like} \\ \hline
$1s^22s^22p^2$ & $1s^2\{2s^22p^2,2p^4\}$ & $1s^2\{2s^22p^2,2p^4,2s2p^23d,2s^23d^2\}$       & $\{8s7p6d5f4g2h\}$        &   340~10$^3$ \\
$1s^22p^4$     & $1s^2\{2s^22p^2,2p^4\}$ & $1s^2\{2s^22p^2,2p^4,2s2p^23d,2s^23d^2\}$       & $\{8s7p6d5f4g2h\}$         &   340~10$^3$ \\
$1s^22s2p^3$   & $1s^22s2p^3$            & $1s^2\{2s2p^3,2p^33d,2s^22p3d,2s2p3d^2\}$       & $\{8s7p6d5f4g2h\}$         & 1~000~10$^3$ \\ \hline
\multicolumn{5}{c}{nitrogen-like} \\ \hline
$1s^22s^22p^3$ & $1s^2\{2s^22p^3,2p^5\}$ & $1s^2\{2s^22p^3,2p^5,2s2p^33d,2s^22p3d^2\}$     & $\{8s7p6d5f4g1h\}$ &   698~10$^3$ \\
$1s^22p^5$     & $1s^2\{2s^22p^3,2p^5\}$ & $1s^2\{2s^22p^3,2p^5,2s2p^33d,2s^22p3d^2\}$     & $\{8s7p6d5f4g1h\}$ &   382~10$^3$ \\
$1s^22s2p^4$   & $1s^22s2p^4$            & $1s^2\{2s2p^4,2p^43d,2s^22p^23d,2s2p^23d^2\}$   & $\{8s7p6d5f4g1h\}$ &   680~10$^3$ \\ \hline
\end{tabular}
\end{table}

\newpage

\section{Results and data evaluation}
In the following subsections, the results of the calculations are presented for the four isoelectronic sequences, and compared with values from the available literature. 

The identification of the levels is based on the $LS$ composition obtained by transforming from $jj$- to $LS$-coupling schemes using the  \textsc{jj2lsj} tool integrated in the new release of \textsc{Grasp2K}~\cite{Jonetal:2013a}. The two first CSFs with weight $\vert c_j \vert^2 \geq 0.1\%$ are also displayed in the level compositions. All the IS electronic parameters reported in the tables have been estimated using  \textsc{ris3} and the correspondence with the $LSJ$ levels has been done using the tool \textsc{ris3\_lsj}~\cite{Nazetal:2013a}.

\subsection{Beryllium-like ions} 
Table 1 gives the total energies, the transition energies, the $\widetilde{K}_{\rm NMS}$ and $\widetilde{K}_{\rm SMS}$ parameters, and the electronic field shift factors $\widetilde {\cal F}$  for levels in beryllium-like ions ($5 \le Z \le 74$) from RCI calculations described in the previous section. 
It must be stressed that the level $2p^2~^1S_0$ of the B II ion required a special attention.  For this excited state, the influence of the triple and quadruple excitations is strong and limiting the population as described in section~4.2 becomes inadequate. The results presented are those of an expansion based on SDTQ-excitations up to $n=7i$. 

Table \ref{tab:Liz_comp} presents a comparison with theoretical and experimental results of Litz\'en \textit{et al.}~\cite{Litetal:98a} for B II, for which 
multiconfiguration Hartree-Fock (MCHF) calculations were performed in a non-relativistic scheme with the program \textsc{atsp2K} \cite{Froetal:2007a}. Expecting a small contribution of orbitals with high  angular momenta, the latter were limited to $l\leq3$.  The $\delta\sigma_{\rm FS}$ values evaluated from (\ref{Line_FS_sm_cm}) and using the experimental nuclear rms charge radius of Angeli~\cite{Ang:2004a}, are reported in the last column. No data are available in the literature, but these numbers illustrate how small the FS is for the B~II ion. An analysis of the 1362~\AA~line profile observed in Hubble spectra of the interstellar medium indicated an IS of $13.7\pm3.5$~ m\AA~for the transition $2s^2~^1S_0-2s2p~^1P^o_1$~\cite{Fedetal:96a} for the pair $^{11}$B-$^{10}$B; converting our value $\delta\sigma_{\rm MS}=0.711$~cm$^{-1}$ gives $\delta\lambda=13.20$ m\AA. As the values obtained by Litz\'en \textit{et al.} are 13~m\AA~and 13.27 m\AA~for experiment and theory respectively, all values are within experimental error bars of the Hubble observation. 
Except for the transition $2s2p~^1P_1^o-2p^2~^1S_0$, values are in good agreement with both experiments and previous theoretical work. The source of the discrepancy has been investigated in details and is attributed to the incomplete correlation model adopted in \cite{Litetal:98a}.

Table \ref{tab:Kor_comp} gives a comparison with the calculations of Korol and Kozlov \cite{KorKoz:2007a}, J\"onsson \textit{et al.}~\cite{Jonetal:99a} and Berengut \textit{et al.}~\cite{Beretal:2006a} for astrophysically important transitions in C III. All values are here expressed in GHz~u. The consistency with the NMS parameter is excellent for all transitions. The SMS parameters are in agreement with the non-relativistic calculations of J\"onsson \textit{et al.} as well as the values of Berengut \textit{et al.} for the transition $1s^22s^2~^1S_0-1s^22s2p~~^3P^o_0$. Note however that our SMS parameter values grow with the $J$ value, in contradiction with Berengut~\textit{et al.}'s results~\cite{Beretal:2006a}. By summing the NMS and SMS contributions for the (13,12) and (14,12) isotopic pairs, the total mass shift values of Berengut~\textit{et al.}, $\delta\nu^{13,12}_{\rm MS}=28.76$~GHz and $\delta\nu^{14,12}_{\rm MS}=53.33$~GHz, are confirmed by our results ($28.86$~GHz and $53.51$~GHz respectively). In their calculations, Berengut~\textit{et al.}, safely neglected the FS that we estimate to a few~MHz. 

Table \ref{tab:Bubin_comp} compares the wavenumber mass shifts calculated by  Bubin \textit{et al.}~\cite{Bubetal:2010a} using explicitly correlated Gaussian functions for the transition $1s^2 2s^2~^1S_0- 1s^2 2p^2~^1S_0$ in C III with our results. 
Considering the high level of correlation included in~\cite{Bubetal:2010a}, the almost perfect agreement between the two sets gives us confidence in our correlation models and computational strategy.
The wavenumber field shifts have been estimated  using~(\ref{Line_FS_sm_cm}) and taking the difference of the nuclear mean-square radii $\delta \langle r^2\rangle^{13,12}=-0.044$ fm$^2$ and $\delta \langle r^2\rangle^{14,13}=0.210$~fm$^2$ from Angeli~\cite{Ang:2004a}. 

We report in table~\ref{tab:King_comp} our  relativistic isotope shift parameters  ($K_{\rm NMS}$ and $K_{\rm SMS}$)
 for the ground state of beryllium-like ions  B~II-Ne~VII and compare them with the non relativistic values obtained by King \textit{et al.}~\cite{Kinetal:2011a} using a standard Hylleraas approach with Slater-type orbital (STO) basis functions, by Komasa {\it et al.}~\cite{Kometal:2002a} 
 using the explicitly correlated Gaussians approach, and by Galvez {\it et al.}~\cite{Galetal:99a} using the Monte Carlo algorithm, starting from explicitly correlated multideterminant wave functions.
 The one-body SMS parameters agree within 0.02\% while the two-body SMS parameters do match within 0.4\% for B~II and 3\% for Ne~VII. For the density operator, the disagreement goes from 1\% for B~II to 4\% for Ne~VII. The comparison illustrates the importance of relativistic effects also for these ions.
 
To conclude this section on Be-like ions, we report in table~\ref{tab:Ort_comp} the $^{40,36}$Ar isotope shift  of the M1 intra-configuration transition $1s^2 2s2p~^3P^o_1$-$^3P^o_2$ in Ar$^{14+}$ that has been measured by Orts {\it et al.}~\cite{Ortetal:2006a} using the electron beam ion trap at the Max-Planck-Institut f\"ur Kernphysik~\cite{Ortetal:2006a}, and estimated theoretically in the same work~\cite{Ortetal:2006a}, using the large-scale configuration interaction Dirac-Fock approach. 
As far as the MS is concerned, the consistency between the theoretical calculations is perfect.
The experimental value is 5.3\% smaller than the theoretical values that lie however within the experimental error bars. 
Our FS value is evaluated using the formula (\ref{Line_FS_sm_cm}) and the difference of  nuclear rms $\delta \langle r^2\rangle^{40,36}=0.2502$~fm$^2$. It also shows a nice consistency with the evaluation of Orts {\it et al.}~\cite{Ortetal:2006a}.

\subsection{Boron-like ions} 
Table~\ref{tab:Blike} displays the total and excitation energies, the $ \widetilde K_{\rm NMS}$ and $ \widetilde K_{\rm SMS}$ parameters as well as the electronic $\widetilde {\cal F}$ field shift  factors  for several levels of B-like ions ($8\leq Z \leq  30$ and $Z=36,42$). 
Table~\ref{tab:Tup_comp} compares the resulting $^{36,40}$Ar$^{13+}$ isotope shift of the  $1s^22s^2 2p~^2P^o_{1/2}~-~^2P^o_{3/2}$ forbidden line with the results of Tupitsyn \textit{et al.}~\cite{Tupetal:2003a} and Orts \textit{et al.}~\cite{Ortetal:2006a} using the large-scale RCI Dirac-Fock method to solve the Dirac-Coulomb-Breit equation.
Their CSFs expansions were generated including ``all single and double excitations and some part of triple excitations''. The nuclear charge distribution is described by a Fermi model and is therefore consistent with the present work. 
The upper part of the table shows the individual contributions of operators $H_{\rm NMS}$ and~$H_{\rm SMS}$ to the wavenumber mass shift. 
A good agreement is observed between the two sets of values, the total wavenumber mass shift values differing by less than 0.8\%.  This example confirms the importance of the relativistic corrections to the recoil operator: the total wavenumber mass shift would indeed be  50\% 
smaller if estimated from the uncorrected form of the mass Hamiltonian $\langle H^1_{\rm NMS} + H^1_{\rm SMS} \rangle$. The large cancellation of the terms involved in the transition isotope shift makes accurate calculations very challenging.  
For the sake of a fair comparison, the  QED contributions ($-0.0006$~cm$^{-1}$) estimated by Orts \textit{et al.}~\cite{Ortetal:2006a} have been substracted from their theoretical estimation.
 The mass shifts values are extremely consistent. Orts \textit{et al.}~\cite{Ortetal:2006a} also measured the transition IS at the Heidelberg electron beam ion trap  with a satisfactory precision and the agreement observation-theory is rather good, as illustrated in 
 table \ref{tab:Tup_comp}.

As an additional check of our RCI values at the neutral end of the isoelectronic sequence, we performed multiconfiguration Hartree-Fock (MCHF) calculations for O~IV. The MCHF calculations are fully variational and the expansions obtained by SD-excitations from large MR sets to an active set of orbitals with $nl \le9h$. Results from these independent calculations are compared with the RCI values in table \ref{tab:B_MCHF_comp}. The two different sets of specific mass shift parameters values differ by less than 1\%, considering both the uncorrected term ($K^1_{\rm SMS}$) or the corrected one ($K_{\rm SMS}$).  The two sets of calculations give specific mass shift parameters that on average differ by less than 0.4\%,  indicating that the uncertainties in the specific mass shift parameters from the present RCI calculations are very small.

\subsection{Carbon-like ions}
\begin{figure}[ht!]
\begin{center}
\includegraphics[width=0.60\textwidth]{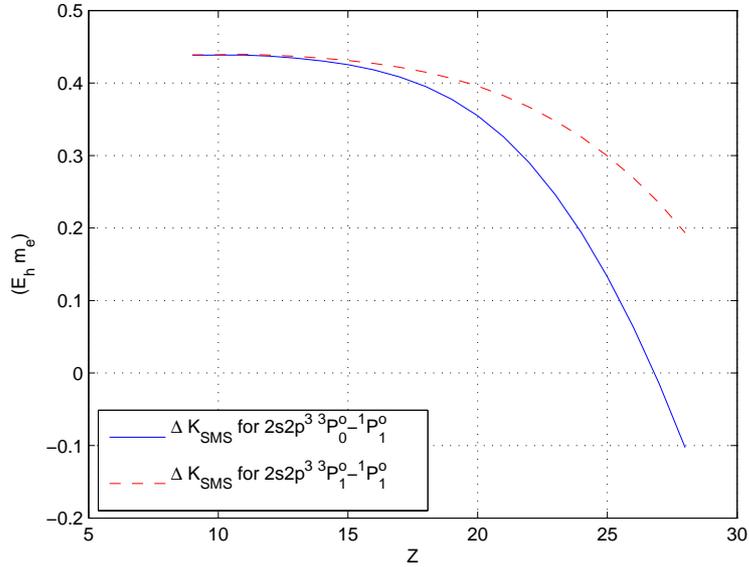}
\end{center}
\caption{Variation of the transition specific mass shift parameters along the isoelectronic sequence  of C-like ions ($7\leq Z \leq28$) for  $2s2p^3~^3P^o_0-~^1P_1^o$ (continuous blue line) and $2s2p^3~^3P^o_1-~^1P_1^o$ (dashed red line). \label{fig:convergence1P3P}}
\end{figure}

Table \ref{tab:Clike} displays the total and excitation energies, the $ \widetilde K_{\rm NMS}$ and $ \widetilde K_{\rm SMS}$ parameters as well as the electronic $\widetilde {\cal F}$ field shift  factors  for several levels of C-like ions ($7\leq Z \leq28$). 
Table \ref{tab:Jon_C} presents the comparison with J\"onsson and Biero\'n's results~\cite{JonBie:2010a} who calculated the uncorrected $K^1_{\rm SMS}$ parameters. For each of the five ions considered in \cite{JonBie:2010a} (N~II, O~III, F~IV, Ne~V and Ti~XVII), the $K^1_{\rm SMS}$ values are compared with each other, and with the corrected $K_{\rm SMS}$ values evaluated in the present work (third column). The small difference between the $K^1_{\rm SMS}$ values is due to differences in the optimization strategy. Furthermore, as generally observed \cite{Gaietal:2011a,Kozetal:2010a,Lietal:2012a}, the relativistic corrections to the level specific mass shift barely reach a third of percent for light ions but reach 2\% for Ti~XVII. J\"onsson and Biero\'n observed that  the SMS parameters for $2s2p^3~^3P^o$ and $2s2p^3~^1P^o$ approach each other for large~$Z$. The present study extends the nuclear charges range considered and, as it can be seen from figure~\ref{fig:convergence1P3P} that plots the differences of the SMS parameter values ($\Delta K_{\rm SMS}$) versus the nuclear charge for  $^3P^o_{1,0}- ^1P^o_{1}$,  the parameters indeed approach each other but this trend is not asymptotic. Some $J$-dependence appears: the sign of $\Delta K_{\rm SMS}$ for $2s2p^3~^3P^o_0$ and $2s2p^3~^1P^o_1$  changes between $Z=27$ and $Z=28$, while for the transition $2s2p^3~^3P^o_1-~^1P_1^o$, the studied nuclear charges range of nuclear charges  does not allow to observe the point where the $\Delta K_{\rm SMS}$ cross the $X$-coordinate axis. Using a third degree polynomial extrapolation, the crossing is predicted to occur between $Z=31$ and $Z=32$. 
 To further validate the present results, independent non relativistic SD-MR-MCHF calculations were performed for a few levels of O~III.
 As it can be seen in table \ref{tab:C_MCHF_comp},  the two sets of values agree very well, with an average difference 0.37\% for $K^1_{SMS}$, as expected for the neutral end of the isoelectronic sequence. 

\subsection{Nitrogen-like ions} 
Table \ref{tab:Nlike} displays the total and excitation energies, the $ \widetilde K_{\rm NMS}$ and $ \widetilde K_{\rm SMS}$ parameters as well as the electronic $\widetilde {\cal F}$ field shift  factors  for several levels for N-like ions ($8\leq Z \leq  30$ and $Z=36,42$). 
To the knowledge of the authors, there are no isotope shift data to compare with along this sequence. To validate the current data independent SD-MR-MCHF calculations were performed for Ne~IV. The comparison is presented in table \ref{tab:N_MCHF_comp}, illustrating the good consistency between the two sets of results. 

\section{Outlook and conclusion}
It is shown that relativistic configuration interaction calculations are capable of providing very accurate electronic isotope shift parameters for few-electron systems. The parameters are of value for interpreting current and future heavy-ion storage ring experiments
and measurements on radioisotopes at ISOLDE to access changes in nuclear charge radii. However, it remains a true challenge to extend the current calculations to larger systems with a core. The fundamental problems are the large and canceling contributions to the mass shift from all core subshells. To obtain accurate results for the electronic mass shift parameters all electrons, even the ones deep down in the core, need to be correlated, leading to massive CSF expansions~\cite{Froetal:98a}. 
One way to handle these problems is to rely on divide and conquer strategies and divide the original large problem into a series of smaller ones for each pair of electrons that should be correlated. Contributions from all electron pairs are then put together at the end in one calculations that describes the full atomic system \cite{Veretal:2013a,Froetal:2013a}. 
Work along these lines are in progress~\cite{Godetal:2013a}.

\ack
Part of this work was supported by the Communaut\'e fran\c caise of Belgium (Action de Recherche Concert\'ee), the Belgian National Fund for Scientific Research (FRFC/IISN Convention) and by the IUAP Belgian State Science Policy (BriX network P7/12).
CN and SV are grateful to the ``Fonds pour la formation ‡ la Recherche dans l'Industrie et dans l'Agriculture'' of Belgium for a PhD grant (Boursier F.R.S.-FNRS). PJ gratefully acknowledges financial support from the Swedish Research Council (VR). PJ and GG acknowledges support from the Visby program of the Swedish Institute.

%%  All sections inside the appendix environment will be appendixes
%%  Subsections function normally in appendixes.

%\noteinproof
%A note added in proof, if there is one, should be the final text before
%the references.
\newpage

%\bibliography{/Users/michelgodefroid/Documents/mrg/atoms}
%\bibliography{atoms}

%%% Please start a new page by uncommenting the next
\newpage
\TableExplanation
\bigskip
\renewcommand{\arraystretch}{1.0}

\section*{Table \ref{tab:Belike}.\label{table_ISBe} 
Total energies (in $E_{\text h}$), excitation energies (in cm$^{-1}$), normal and specific mass shift $\widetilde{K}$ parameters (in GHz u), $\widetilde {\cal F}$ field shift factors (in GHz/fm$^2$) for levels in the beryllium isoelectronic sequence ($5\leq Z\leq 74$). For each of the many-electron wave functions, the leading components are given in the $LSJ$ coupling scheme. The number in square brackets is the power of 10.}
\begin{tabular*}{0.95\textwidth}{@{}@{\extracolsep{\fill}}lp{5.5in}@{}}
Level composition in $LSJ$ coupling & Leading components expressed within the $LSJ$ coupling scheme \\
								   & for each calculated level. \\
Energy          & The calculated total energy in atomic units ($E_{\text h}$).\\
$\Delta E$      & Calculated energy relative to the ground state (in cm$^{-1}$).\\
$\widetilde K_{\rm NMS} =  K_{\rm NMS}/h $   & The normal mass shift parameter in GHz~u (see eq.(\ref{eq:NMS})).\\
$\widetilde K_{\rm SMS} = K_{\rm SMS}/h  $   & The specific mass shift parameter in GHz~u (see eq.(\ref{eq:SMS})).\\
$\widetilde {\cal F}$         & The electronic field shift factor in GHz/fm$^2$ (see eq.(\ref{level_electronic_factor})). \\
\end{tabular*}

\section*{Table \ref{tab:Liz_comp}.\label{table_IS_Liz} Comparison of wavenumber isotope shifts $\delta\sigma$ (in cm$^{-1}$)  for several transitions in B II between the isotopes $^{11}$B and $~^{10}$B.}
\begin{tabular*}{0.95\textwidth}{@{}@{\extracolsep{\fill}}lp{5.5in}@{}}
Exp. $\delta\sigma_{\rm IS}$&Experimental transition IS (in cm$^{-1}$)\\
$\delta\sigma_{\rm MS}$&Transition mass shift (in cm$^{-1}$).\\
$\delta\sigma_{\rm FS}$&Transition field shift  (in cm$^{-1}$) (see eq.(\ref{Line_FS_sm_cm})).\\\\
\end{tabular*}

\section*{Table \ref{tab:Kor_comp}. Comparison of the $\Delta \widetilde K_{\rm NMS}$ and $\Delta \widetilde K_{\rm SMS}$ parameters (in~GHz~u) with previous theoretical works for several transitions in~C~III.}
\begin{tabular*}{0.95\textwidth}{@{}@{\extracolsep{\fill}}lp{5.5in}@{}}
$\Delta \widetilde K_{\rm NMS}$&Transition normal mass shift (NMS) parameters (in GHz u). \\
$\Delta \widetilde K_{\rm SMS}$&Transition specific mass shift (SMS) parameters (in GHz u).\\
\end{tabular*}
\section*{Table \ref{tab:Bubin_comp}. Wavenumber mass shift $\delta\sigma_{\rm MS}$ and field shift $\delta\sigma_{\rm FS}$ (in cm$^{-1}$) for the transition $1s^2 2s^2~^1S_0- 1s^2 2p^2~^1S_0$ in C$^{2+}$ ion for the isotope pairs $^{13}$C-$^{12}$C and $^{14}$C-$^{13}$C.}
\begin{tabular*}{0.95\textwidth}{@{}@{\extracolsep{\fill}}lp{5.5in}@{}}
$\delta\sigma_{\rm MS}$&Transition mass shift (in cm$^{-1}$).  \\
$\delta\sigma_{\rm FS}$ &Transition field shift (in cm$^{-1}$).\\
\end{tabular*}

\section*{Table \ref{tab:King_comp}. Comparative table of the expectation values of our isotope shift parameters (in a.u.) with calculations based on Hylleraas wave functions for some ions at the neutral end of the Be-like sequence.}
\begin{tabular*}{0.95\textwidth}{@{}@{\extracolsep{\fill}}lp{5.5in}@{}}
$K_{\rm NMS}$ & Level normal mass shift parameter (see eq.(\ref{eq:NMS})) calculated with the one-body part of (\ref{eq:H_MS}). \\
$\langle -\frac{1}{2}\nabla_i^2\rangle$ & Non relativistic kinetic energy.\\
$K_{\rm SMS}$ & Level specific mass shift parameter (see eq.(\ref{eq:SMS})) calculated with the two-body part of (\ref{eq:H_MS}).  \\
$\rho^e({\bf 0})$& Total probability density at the origin estimated from eq.(\ref{el_dens_dif_1}). \\
$\langle \delta({\bf r}_i) \rangle$& Non relativistic total probability density at the origin (nuclear point charge). \\
\end{tabular*}

\section*{Table \ref{tab:Ort_comp}. Wavenumber mass shift $\delta\sigma_{\rm MS}$ and field shift $\delta\sigma_{\rm FS}$ (in cm$^{-1}$) for the $1s^2 2s2p~^3P^o_1-~^3P^o_2$ in Ar$^{14+}$  for the isotope pair $^{40,36}$Ar.}
\begin{tabular*}{0.95\textwidth}{@{}@{\extracolsep{\fill}}lp{5.5in}@{}}
&Same as for Table \ref{tab:Bubin_comp}.
\end{tabular*}

\section*{Table \ref{tab:Blike}.\label{table_ISB} 
Total energies (in $E_{\text h}$), excitation energies (in cm$^{-1}$), normal and specific mass shift $\widetilde K$ parameters (in GHz u), $\widetilde {\cal F}$ field shift factors (in GHz/fm$^2$) for levels in the boron isoelectronic sequence ($8\leq Z \leq  30$ and $Z=36,42$). For each of the many-electron wave functions, the leading components are given in the $LSJ$ coupling scheme.  The number in square brackets is the power of 10.}
\begin{tabular*}{0.95\textwidth}{@{}@{\extracolsep{\fill}}lp{5.5in}@{}}
&Same as for Table \ref{tab:Belike}.
\end{tabular*}
\section*{Table \ref{tab:Tup_comp}. Upper part: individual contributions to the wavenumber mass shift $\delta \sigma_{\rm MS}$ of the forbidden transition $1s^2 2s^2~2p~^2 P^o_{1/2}-~^2P^o_{3/2}$ in boron-like $^{36,40}$Ar$^{13+}$. Lower part: wavenumber mass, field and total shifts.  (All numbers in cm$^{-1}$).}
\begin{tabular*}{0.95\textwidth}{@{}@{\extracolsep{\fill}}lp{5.5in}@{}}
$\langle H^i_{\rm NMS}\rangle$~$(i=1,2,3)$ &  Normal mass shift (one-body) contributions from the three terms of eq.(\ref{eq:H_MS}). \\
$\langle H^i_{\rm SMS}\rangle$~$(i=1,2,3)$ &  Specific mass shift (two-body) contributions from the three terms of eq.(\ref{eq:H_MS}). \\
$\delta\sigma_{\rm MS}$&Transition mass shift (in cm$^{-1}$).  \\
$\delta\sigma_{\rm FS}$ &Transition field shift (in cm$^{-1}$).\\
$\delta\sigma$ &Total transition isotope shift (in cm$^{-1}$). \\
\end{tabular*}

\section*{Table \ref{tab:B_MCHF_comp}. Comparison of MCHF and MCDHF specific mass shift parameters for O~IV (all values in atomic units $E_{\rm h}m_e$).}
\begin{tabular*}{0.95\textwidth}{@{}@{\extracolsep{\fill}}lp{5.5in}@{}}
MCHF $K^1_{\rm SMS}$             & Non relativistic specific mass shift parameter. \\
MCDHF $K^1_{\rm SMS}$             & Relativistic specific mass shift parameter calculated with $\frac{1}{2}
 \langle  \sum_{i \neq j} ^{N} \; \left( {\bf  p}_{i} \cdot {\bf  p}_{j} \right) \rangle $. \\
MCDHF $K_{\rm SMS}$             & Relativistic specific mass shift parameter calculated with the two-body part of the relativistic recoil operator (\ref{eq:H_MS}). 
\end{tabular*}

\section*{Table \ref{tab:Clike}.\label{table_ISC} Total energies (in $E_{\text h}$), excitation energies (in cm$^{-1}$), normal and specific mass shift $\widetilde{K}$ parameters (in GHz u), $\widetilde {\cal F}$ field shift factors (in GHz/fm$^2$) for levels in the carbon isoelectronic sequence ($7\leq Z \leq28$). For each of the many-electron wave functions, the leading components are given in the $LSJ$ coupling scheme. The number in square brackets is the power of 10.}
\begin{tabular*}{0.95\textwidth}{@{}@{\extracolsep{\fill}}lp{5.5in}@{}}
&Same as for Table \ref{tab:Belike}.
\end{tabular*}
\section*{Table \ref{tab:Jon_C}.\label{table_ISjon_C} Comparison of corrected ($K_{\rm SMS}$)  and  uncorrected ($K^1_{\rm SMS}$) specific shift parameters (in $E_{\rm h} m_e $) for several states of N~II, O~III, F~IV, Ne~V and Ti~XVII.}
\begin{tabular*}{0.95\textwidth}{@{}@{\extracolsep{\fill}}lp{5.5in}@{}}
$K^1_{\rm SMS}$             & Relativistic specific mass shift parameter calculated with $\frac{1}{2}
 \langle  \sum_{i \neq j} ^{N} \; \left( {\bf  p}_{i} \cdot {\bf  p}_{j} \right) \rangle $. \\
$K_{\rm SMS}$             & Relativistic specific mass shift parameter calculated with the two-body part of the relativistic recoil operator (\ref{eq:H_MS}). 
\end{tabular*}
\section*{Table \ref{tab:C_MCHF_comp}. Comparison of MCHF and MCDHF specific mass shift parameters for O~III (in atomic units $E_{\rm h}m_e$).}
\begin{tabular*}{0.95\textwidth}{@{}@{\extracolsep{\fill}}lp{5.5in}@{}}
&Same as for Table \ref{tab:B_MCHF_comp}.
\end{tabular*}
\section*{Table \ref{tab:Nlike}. \label{table_ISN}Total energies (in $E_{\text h}$), excitation energies (in cm$^{-1}$), normal and specific mass shift $\widetilde K$ parameters (in GHz u), $\widetilde {\cal F}$ field shift factors (in GHz/fm$^2$) for levels in the nitrogen isoelectronic sequence ($7\leq Z\leq36$ and $Z=42,74$). For each of the many-electron wave functions, the leading components are given in the $LSJ$ coupling scheme. The number in square brackets is the power of 10.}
\begin{tabular*}{0.95\textwidth}{@{}@{\extracolsep{\fill}}lp{5.5in}@{}}
&Same as for Table \ref{tab:Belike}.
\end{tabular*}
\section*{Table \ref{tab:N_MCHF_comp}. Comparison of MCHF and MCDHF specific mass shift parameters for Ne~IV (in atomic units $E_{\rm h}m_e$).}
\begin{tabular*}{0.95\textwidth}{@{}@{\extracolsep{\fill}}lp{5.5in}@{}}
&Same as for Table \ref{tab:B_MCHF_comp}.
\end{tabular*}

\newpage

\datatables % This command is necessary to get the table names in toc

%% One-page data tables are also best formatted using the longtable
%% environment:
%\begin{longtable}{c}
%\caption{This is the First Data Table}\\
%\endhead\\
%\end{longtable}

%% If the table is to span over the whole text width, we set:

\setlength{\LTleft}{0pt}
\setlength{\LTright}{0pt} 

% To avoid ``Overfull \hboxes...'' decrease the intercolumn spacing:

\setlength{\tabcolsep}{0.5\tabcolsep}

\renewcommand{\arraystretch}{1.0}

\footnotesize % we need to squeeze the font size a lot!

\newpage

\renewcommand{\arraystretch}{1.0}

\footnotesize
\begin{longtable}{@{\extracolsep\fill}llllcrrrrc@{}}
\caption{Total energies (in $E_{\text h}$), excitation energies (in cm$^{-1}$), normal and specific mass shift $\widetilde{K}$ parameters (in GHz u), $\widetilde {\cal F}$ field shift factors (in GHz/fm$^2$) for levels in the beryllium isoelectronic sequence ($5\leq Z\leq 74$). For each of the many-electron wave functions, the leading components are given in the $LSJ$ coupling scheme.  The number in square brackets is the power of 10. See page\ \pageref{table_ISBe} for Explanation of Tables. \label{tab:Belike}}\\ [-.4cm]
%\multicolumn{4}{c}{Level composition in $LSJ$ coupling} & \multicolumn{1}{c}{$J$} & \multicolumn{1}{c}{Energy ($E_{\text h}$)} &\multicolumn{1}{c}{$\Delta E$ (cm$^{-1}$)}&  \multicolumn{1}{c}{$K_{\rm NMS}$ ($E_{\text h} m_e$)} &  \multicolumn{1}{c}{$K_{\rm SMS}$ ($E_{\text h} m_e$)}  &  \multicolumn{1}{c}{Density ($a_0^{-3}$)} \\
\multicolumn{4}{c}{Level composition in $LSJ$ coupling} & \multicolumn{1}{c}{$J$} & \multicolumn{1}{c}{Energy ($E_{\text h}$)} &\multicolumn{1}{c}{$\Delta E$ (cm$^{-1}$)}&  \multicolumn{1}{c}{$\widetilde K_{\rm NMS}$ (GHz u )} &  \multicolumn{1}{c}{$\widetilde K_{\rm SMS}$ (GHz u )}  &  \multicolumn{1}{c}{$\widetilde {\cal F}$~(GHz/fm$^{2}$)} \\                                                                        
\endfirsthead
\caption[]{(continued)} \\ [-.4cm]
\multicolumn{4}{c}{Level composition in $LSJ$ coupling} & \multicolumn{1}{c}{$J$} & \multicolumn{1}{c}{Energy ($E_{\text h}$)} &\multicolumn{1}{c}{$\Delta E$ (cm$^{-1}$)}&  \multicolumn{1}{c}{$\widetilde K_{\rm NMS}$ (GHz u )} &  \multicolumn{1}{c}{$\widetilde K_{\rm SMS}$ (GHz u )}  &  \multicolumn{1}{c}{$\widetilde {\cal F}$~(GHz/fm$^{2}$)} \\ 
\hline
\endhead
\noalign{\vskip3pt}
\multicolumn{6}{r}{\itshape Continued\dots}\\
\endfoot
\noalign{\vskip3pt}\hline
\endlastfoot
\hline&&&& \multicolumn{3}{c}{B~II} \\ \hline
$1s^22s^2$ $^1S$ &(92\%) &+ $1s^22p^2$ $^1S$ &(7\%) &0 &$ -24.3534464 $&$0.00 $&$8.7862981[4]$& $2.1587130[3]$& 1.803445  \\ 
$1s^22s2p$ $^3P^o$ &(99\%) &&&0 &$ -24.1836233 $&$37271.86 $&$8.7249235[4]$& $3.1895258[2]$& 1.763953  \\ 
&(99\%) &&&1 &$ -24.1835957 $&$37277.92 $&$8.7249488[4]$& $3.1855676[2]$& 1.763955  \\ 
&(99\%) &&&2 &$ -24.1835226 $&$37293.97 $&$8.7249802[4]$& $3.1780709[2]$& 1.763958  \\ 
$1s^22s2p$ $^1P^o$ &(96\%) &+ $1s^22p3d$ $^1P^o$ &(3\%) &1 &$ -24.0185813 $&$73494.38 $&$8.6679944[4]$& $9.8372368[2]$& 1.759018  \\ 
$1s^22p^2$ $^3P$ &(99\%) &&&0 &$ -23.9028711 $&$98889.85 $&$8.6233220[4]$& $-1.1484849[3]$& 1.715645  \\ 
&(99\%) &&&1 &$ -23.9028320 $&$98898.43 $&$8.6233347[4]$& $-1.1488285[3]$& 1.715646  \\
\end{longtable}
%\clearpage
\begin{table}[htdp]
\begin{center}\caption{\label{tab:Liz_comp}Comparison with other works (experiment and theory) of wavenumber isotope shifts $\delta\sigma$ (in cm$^{-1}$)  for several transitions in B II between the isotopes $^{11}$B and $~^{10}$B.} 
\begin{tabular}{@{}l|lcccc@{}} 
%\multicolumn{2}{c}{ }\\
\toprule
%&\multicolumn{4}{c}{$\delta \sigma$}\\
%\multicolumn{2}{c}{Litz\'en \textit{et al.} \cite{Litetal:98a}}
&\multicolumn{1}{c}{Exp.}&Theory &&\multicolumn{2}{c}{This work}\\
\cline{5-6}
&\multicolumn{1}{c}{$\delta\sigma_{\rm IS}$}&\multicolumn{1}{c}{$\delta\sigma_{\rm MS}$}
&&$\delta\sigma_{\rm MS}$&$\delta\sigma_{\rm FS}$\\
%\multicolumn{1}{c}{Total}\\
\midrule 
\\
$2s^2~^1S_0-2s2p~^1P^o_1$&0.70$^a$&0.7145$^a$	&&0.711&$\phantom{-}$0.00016\\
						&0.74(19)$^b$&0.7146$^c$&&&\\
						&&0.7165$^d$&&&\\
$2s2p~^3P^o_1-2p^2~^3P_2$&0.80(4)$^a$&0.7485$^a$&&0.749&$-0.00017$\\\\
%$2s2p~^3P^o_?-2p^2~^3P_?$&&\\
$2s2p~^1P^o_1-2p^2~^1S_0$&0.419$^a$&0.4263$^a$&&0.478&$-0.00010$\\
$2s2p~^1P^o_1-2p^2~^1D_2$&0.887$^a$&0.8870$^a$&&0.890&$-0.00014$\\
\\
\bottomrule
\multicolumn{3}{l}{$^a$ Litz\'en \textit{et al.} \cite{Litetal:98a}}\\
\multicolumn{5}{l}{$^b$ Astronomical observations from Federman \textit{et al.}~\cite{Fedetal:96a}}\\
\multicolumn{5}{l}{$^c$ J\"onsson \textit{et al.} \cite{Jonetal:99a}}\\
\multicolumn{6}{l}{$^d$ MCHF calculations of J\"onsson \textit{et al.}~\cite{Jonetal:94a} - no limits on orbital quantum numbers.}\\
\end{tabular}\end{center}
\end{table}

\begin{table}[htdp]
\centering
\caption{\label{tab:Kor_comp}Comparison of the $\Delta \widetilde K_{\rm NMS}$ and $\Delta \widetilde K_{\rm SMS}$ parameters (in~GHz~u) with previous theoretical works for several transitions in~C~III.} 
\begin{tabular}{@{}lc|cccllc} 
\toprule
%&\multicolumn{6}{c}{C III}\\
&&\multicolumn{2}{c}{$\Delta \widetilde K_{\rm NMS}$}&&\multicolumn{2}{c}{$\Delta \widetilde K_{\rm SMS}$}\\
\cline{3-4}\cline{6-7}
&~~~~~&Other works&This work&&Other works&This work\\
\toprule
&&&&&&\\
$2s^2~^1S-2s2p~^3P^o$& & &&&$-3475^c$&\\
\multicolumn{2}{r|}{$2s^2~^1S_0-2s2p~^3P^o_0$}&$-860.3^a$&$-860.1$%&$-3475^c$
&&$-3453(7)^a$	&$-3474$\\
&&&&&$-3473^b$\\
\multicolumn{2}{r|}{$2s^2~^1S_0-2s2p~^3P^o_1$}&$-860.2^a$&$-859.8$&&$-3454(7)^a$	&$-3475$\\
&&&&&$-3472^b$\\
\multicolumn{2}{r|}{$2s^2~^1S_0-2s2p~^3P^o_2$}&$-860.1^a$&$-859.4$&&$-3454(7)^a$	&$-3477$\\
&&&&&$-3468^b$\\
&&&&\\
$2s^2~^1S-2s2p~^1P^o$&&&&&$-2817^c$\\
\multicolumn{2}{r|}{$2s^2~^1S_0-2s2p~^1P^o_1$}&$-1677.3^a$&$-1660$&&$-2774(80)^a$	&$-2826$\\
&&&&&$-2790^b$\\
\bottomrule
\multicolumn{3}{l}{$^a$Korol and Kozlov \cite{KorKoz:2007a}}\\
\multicolumn{3}{l}{$^b$Berengut \textit{et al.} \cite{Beretal:2006a}}\\
\multicolumn{3}{l}{$^c$J\"onsson \textit{et al.} \cite{Jonetal:99a}}\\
\end{tabular}
\end{table}
%\newpage

\begin{table}[htdp]
\begin{center}
\caption{\label{tab:Bubin_comp}Wavenumber mass shift $\delta\sigma_{\rm MS}$ and field shift $\delta\sigma_{\rm FS}$ (in cm$^{-1}$) for the transition $1s^2 2s^2~^1S_0- 1s^2 2p^2~^1S_0$ in C$^{2+}$ ion for the isotope pairs $^{13}$C-$^{12}$C and $^{14}$C-$^{13}$C.}
\begin{tabular}{@{}l|ll|ll} 
\toprule 
&\multicolumn{1}{c}{$\delta\sigma^{13,12}_{\rm MS}$}&\multicolumn{1}{c|}{$\delta\sigma^{13,12}_{\rm FS}$}&
\multicolumn{1}{c}{$\delta\sigma^{14,13}_{\rm MS}$}&\multicolumn{1}{c}{$\delta\sigma^{14,13}_{\rm FS}$}\\
\midrule 
This work&  1.750& 0.00032&1.493&$-0.00151$\\
Bubin \textit{et al.}~\cite{Bubetal:2010a}  & 1.755& &1.498&\\
\bottomrule 
\end{tabular}
\end{center}
\end{table}
%mrg
\begin{table}[htdp]
\centering
\caption{\label{tab:King_comp}Comparison of the electronic isotope shift parameters with previous non-relativistic works for the ground states of beryllium-like B~II - Ne~VII.} 
\begin{tabular}{@{}r|lllllllllllllllllllllll} 
\toprule
%&\multicolumn{6}{c}{C III}\\
&\multicolumn{2}{c}{B II}&&\multicolumn{2}{c}{C III}&&\multicolumn{2}{c}{N IV}&&\\
\cline{2-3}\cline{5-6}\cline{8-10}
%&&&&&&\\
%Level energy&$-24.3534464^a$ &&&$-36.5465996^a$&&&$-51.2469981^a$\\
%&$-24.34861^b$&&&$-36.53458^b$&&&$-51.22230^b$\\
%&$-24.348884446(35)^c$&&&$-36.534852338(35)^d$ &&&$-51.2227083^e$&\\\\
$K_{\rm NMS}$&$\phantom{-}24.3423^a$&&&$\phantom{-}36.5247^a$&&&$\phantom{-}51.2080^a$\\
$\langle -\frac{1}{2}\nabla_i^2\rangle$&$\phantom{-}24.34878^b$&&&$\phantom{-}36.53446^b$&&&$\phantom{-}51.22221^b$\\\\
$K_{\rm SMS}$&$\phantom{-}0.5981^a$&&&$\phantom{-}0.7181^a$&&&$\phantom{-}0.8233^a$\\
$-\langle \nabla_i\cdot\nabla_j\rangle$&$\phantom{-}0.59554(40)^b$&&&$\phantom{-}0.71314(50)^b$&&&$\phantom{-}0.81569(50)^b$\\
&$\phantom{-}0.595140^c$&&&$\phantom{-}0.713671^c$&&&$\phantom{-}0.816044^c$\\\\
\end{tabular}
\end{table}
%mrg
\begin{table}[htdp]
\begin{center}
\caption{\label{tab:Ort_comp}Wavenumber mass shift $\delta\sigma_{\rm MS}$ and field shift $\delta\sigma_{\rm FS}$ (in cm$^{-1}$) for the $1s^2 2s2p~^3P^o_1-~^3P^o_2$ in Ar$^{14+}$  for the isotope pair $^{40,36}$Ar.}
\begin{tabular}{@{}ll|ll} 
\toprule
 & &\multicolumn{1}{c}{$\delta\sigma_{\rm MS}^{40,36}$}&\multicolumn{1}{c}{$\delta\sigma_{\rm FS}^{40,36}$}\\
\midrule 
Theory & this work& 0.0359&$-0.00012$\\
& Orts \textit{et al.} \cite{Ortetal:2006a}  &0.0359%(14)%0.0345+0.0002-0.0015-0.0001\\
&$-0.0001$\\
&&& \\
Observation & Orts \textit{et al.} \cite{Ortetal:2006a}  &0.0340(28)&\\
\bottomrule
\end{tabular}\end{center}
\end{table}

\newpage
\renewcommand{\arraystretch}{1.0}
\footnotesize
% !TEX root = adndt_IS_131121.tex
%
\begin{longtable}{@{\extracolsep\fill}llllllrccc@{}}
\caption{Total energies (in $E_{\text h}$), excitation energies (in cm$^{-1}$), normal and specific mass shift $\widetilde K$ parameters (in GHz u), $\widetilde {\cal F}$ field shift factors (in GHz/fm$^2$) for levels in the boron isoelectronic sequence ($8\leq Z \leq  30$ and $Z=36,42$). For each of the many-electron wave functions, the leading components are given in the $LSJ$ coupling scheme. The number in square brackets is the power of 10. See page\ \pageref{table_ISB} for Explanation of Tables. \label{tab:Blike}}\\ [-.4cm]
\multicolumn{4}{c}{Level composition in $LSJ$ coupling} & \multicolumn{1}{c}{$J$} & \multicolumn{1}{c}{Energy ($E_{\text h}$)} &\multicolumn{1}{c}{$\Delta E$ (cm$^{-1}$)}&  \multicolumn{1}{c}{$\widetilde K_{\rm NMS}$ (GHz u )} &  \multicolumn{1}{c}{$\widetilde K_{\rm SMS}$ (GHz u )}  &  \multicolumn{1}{c}{$\widetilde {\cal F}$~(GHz/fm$^{2}$)} \\                                                                       
\endfirsthead
\caption[]{(continued)} \\ [-.4cm]
\multicolumn{4}{c}{Level composition in $LSJ$ coupling} & \multicolumn{1}{c}{$J$} & \multicolumn{1}{c}{Energy ($E_{\text h}$)} &\multicolumn{1}{c}{$\Delta E$ (cm$^{-1}$)}&  \multicolumn{1}{c}{$\widetilde K_{\rm NMS}$ (GHz u )} &  \multicolumn{1}{c}{$\widetilde K_{\rm SMS}$ (GHz u )}  &  \multicolumn{1}{c}{$\widetilde {\cal F}$~(GHz/fm$^{2}$)} \\
%\multicolumn{4}{c}{Level composition in $LSJ$ coupling} & \multicolumn{1}{c}{$J$} & \multicolumn{1}{c}{Energy ($E_{\text h}$)} &\multicolumn{1}{c}{$\Delta E$ (cm$^{-1}$)}&  \multicolumn{1}{c}{$K_{\rm NMS}$ ($E_{\text h} m_e$)} &  \multicolumn{1}{c}{$K_{\rm SMS}$ ($E_{\text h} m_e$)}  &  \multicolumn{1}{c}{Density ($a_0^{-3}$)}\\
\hline
\endhead
\noalign{\vskip3pt}
\multicolumn{6}{r}{\itshape Continued\dots}\\
\endfoot
\noalign{\vskip3pt}\hline
\endlastfoot
\hline&&&& \multicolumn{3}{c}{O~IV} \\ \hline
$1s^22s^22p$ $^2P^o$ &(95\%) &+ $1s^22p^3$ $^2P^o$ &(4\%) &1/2 &$ -71.3006468 $&$0.00 $&$2.5709821[5]$& $-3.8411303[3]$& 1.283334[1] \\ 
&(95\%) &+ $1s^22p^3$ $^2P^o$ &(4\%) &3/2 &$ -71.2988722 $&$389.49 $&$2.5709779[5]$& $-3.8498436[3]$& 1.283339[1] \\ 
$1s^22s2p^2$ $^4P$ &(100\%) &&&1/2 &$ -70.9755351 $&$71353.76 $&$2.5592579[5]$& $-1.1295016[4]$& 1.243923[1] \\ 
&(100\%) &&&3/2 &$ -70.9749349 $&$71485.49 $&$2.5592561[5]$& $-1.1298419[4]$& 1.243924[1] \\ 
&(100\%) &&&5/2 &$ -70.9740914 $&$71670.63 $&$2.5592683[5]$& $-1.1305008[4]$& 1.243926[1] \\ 
$1s^22s2p^2$ $^2D$ &(98\%) &+ $1s^22s^23d$ $^2D$ &(1\%) &5/2 &$ -70.7218546 $&$127030.20 $&$2.5504401[5]$& $-1.0975396[4]$& 1.242509[1] \\ 
&(98\%) &+ $1s^22s^23d$ $^2D$ &(1\%) &3/2 &$ -70.7217904 $&$127044.30 $&$2.5504327[5]$& $-1.0973347[4]$& 1.242510[1] \\
\end{longtable}
%* the composition of this level is: 15$\%$ $2p^3~^2P^o$ + 13$\%$ $2p^3~^4S^o$ + 5$\%$ $2p^3~^2D^o$
\begin{table}[htdp]
\centering
%\begin{footnotesize}
\caption{\label{tab:Tup_comp}Upper part: individual contributions to the wavenumber mass shift $\delta \sigma_{\rm MS}$ of the forbidden transition $1s^2 2s^2~2p~^2 P^o_{1/2}-~^2P^o_{3/2}$ in boron-like $^{36,40}$Ar$^{13+}$. Lower part: wavenumber mass, field and total shifts. All numbers in cm$^{-1}$. } 
\begin{tabular}{llcccc@{}} 
\toprule
&&\multicolumn{4}{c}{$\delta \sigma_{\rm MS}$}\\
\cline{3-6}
& &\multicolumn{1}{c}{$\langle H^1_{\rm NMS}\rangle$}   
&\multicolumn{1}{c}{$\langle H^2_{\rm NMS}+ H^3_{\rm NMS}\rangle$} 
&\multicolumn{1}{c}{$\langle H^1_{\rm SMS}\rangle$}   
&\multicolumn{1}{c}{$\langle H^2_{\rm SMS}+ H^3_{\rm SMS}\rangle$}\\
\midrule 
%&&&&& \\
\multicolumn{1}{l}{Theory} & \multicolumn{1}{l}{This work}                                     &$0.1054$		&$-0.0821$	&$-0.0745$	&$0.1150$	\\
                           & \multicolumn{1}{l}{Tupitsyn \textit{et al.}~\cite{Tupetal:2003a}}	&$0.1053$		&$-0.0822$	&$-0.0742$	&$0.1151$\\
\bottomrule
& & \multicolumn{1}{c}{$\delta \sigma_{\rm MS}$} & \multicolumn{1}{c}{$\delta \sigma_{\rm FS}$} & \multicolumn{1}{c}{$\delta \sigma$}\\
\midrule
\multicolumn{1}{l}{Theory} &  \multicolumn{1}{l}{This work} 	& 0.0639& \multicolumn{1}{c}{$-0.00053$} & 0.0633\\
 & \multicolumn{1}{l}{Tupitsyn  \textit{et al.}~\cite{Tupetal:2003a}}&0.0640					\\
& Orts \textit{et al.} \cite{Ortetal:2006a} 		& 0.0640$^a$ &\multicolumn{1}{c}{$-0.0005$} & 0.0635\\
&&&&\\
\multicolumn{1}{l}{Observed} & \multicolumn{1}{l}{Orts \textit{et al.} \cite{Ortetal:2006a}}  	& & &\multicolumn{1}{l}{0.0640(31)} & \\
\bottomrule
\multicolumn{6}{l}{$^a$ the QED contributions ($-0.0006$~cm$^{-1}$) have been substracted from their estimation for a fair comparison.}
\end{tabular}
%\end{footnotesize}
\end{table}

\begin{table}[htdp]
\centering
%\caption{\label{tab:C_MCHF_comp} Comparison of MCHF and MCDHF specific mass shift parameters for O~III (in atomic units $E_hm_e$).} 
\caption{\label{tab:B_MCHF_comp} Comparison of MCHF and MCDHF specific mass shift parameters for O~IV (in atomic units $E_{\rm h} m_e$).} 
\begin{tabular}{lcccccc@{}} 
\toprule
 &MCHF&&\multicolumn{3}{c}{MCDHF}\\
 \cline{4-6}
&\multicolumn{1}{c}{$K_{\rm SMS}^1$}&&$J$&$K^1_{\rm SMS}$&$K_{\rm SMS}$\\
\midrule
$2s^22p~^2P^o$  	&  $-1.07928$	&&1/2&$-1.0703$&$-1.0642$\\
						&	&&3/2&$-1.0654$&$-1.0666$\\
$2s2p^2~^4P$   	& $-3.14595$	&&1/2&$-3.1404$&$-3.1293$\\
						&	&&3/2&$-3.1388$&$-3.1302$\\
						&	&&5/2&$-3.1365$&$-3.1320$\\
$2s2p^2~^2D$    	&$-3.05072$	&&5/2&$-3.0476$&$-3.0407$\\
						&	&&3/2&$-3.0472$&$-3.0401$\\
$2s2p^2~^2P$    	&$-2.75703$	&&1/2&$-2.7544$& $-2.7431$\\
    						&	&&3/2&$-2.7511$& $-2.7448$\\
$2s2p^2~^2S$   	&$-2.6815$	&&1/2&$-2.6877$&$-2.6809$\\
$2p^3~^4S^o$   	&$-4.8515$	&&3/2&$-4.8454$&$-4.8315$\\
$2p^3~^2D^o$  	&$-5.0423$	&&5/2&$-5.0427$&$-5.0296$\\
						&	&&3/2&$-5.0425$&$-5.0292$\\
$2p^3~^2P^o$   	&$-4.5694$	&&1/2&$-4.5796$&$-4.5667$\\
						&	&&3/2&$-4.5794$&$-4.5671$\\
\bottomrule
\end{tabular}
\end{table}

\newpage
\renewcommand{\arraystretch}{1.0}
\footnotesize
% !TEX root = adndt_IS_131121.tex
\begin{longtable}{@{\extracolsep\fill}llllllrccc@{}}
\caption{\label{tab:Clike}Total energies (in $E_{\text h}$), excitation energies (in cm$^{-1}$), normal and specific mass shift $\widetilde{K}$ parameters (in GHz u), $\widetilde {\cal F}$ field shift factors (in GHz/fm$^2$) for levels in the carbon isoelectronic sequence ($7\leq Z \leq28$). For each of the many-electron wave functions, the leading components are given in the $LSJ$ coupling scheme. The number in square brackets is the power of 10. See page\ \pageref{table_ISC} for Explanation of Tables.}\\ [-.4cm]
\multicolumn{4}{c}{Level composition in $LSJ$ coupling} & \multicolumn{1}{c}{$J$} & \multicolumn{1}{c}{Energy ($E_{\text h}$)} &\multicolumn{1}{c}{$\Delta E$ (cm$^{-1}$)}&  \multicolumn{1}{c}{$\widetilde K_{\rm NMS}$ (GHz u )} &  \multicolumn{1}{c}{$\widetilde K_{\rm SMS}$ (GHz u )}  &  \multicolumn{1}{c}{$\widetilde {\cal F}$~(GHz/fm$^{2})$} \\                                                                  
\endfirsthead
\caption[]{(continued)} \\ [-.4cm]
\multicolumn{4}{c}{Level composition in $LSJ$ coupling} & \multicolumn{1}{c}{$J$} & \multicolumn{1}{c}{Energy ($E_{\text h}$)} &\multicolumn{1}{c}{$\Delta E$ (cm$^{-1}$)}&  \multicolumn{1}{c}{$\widetilde K_{\rm NMS}$ (GHz u )} &  \multicolumn{1}{c}{$\widetilde K_{\rm SMS}$ (GHz u )}  &  \multicolumn{1}{c}{$\widetilde {\cal F}$~(GHz/fm$^{2})$} \\
\hline
\endhead
\noalign{\vskip3pt}
\multicolumn{6}{r}{\itshape Continued\dots}\\
\endfoot
\noalign{\vskip3pt}\hline
\endlastfoot
\hline&&&& \multicolumn{3}{c}{N~II} \\ \hline
$1s^22s^22p^2$ $^3P$ &(96\%) &+ $1s^22p^4$ $^3P$ &(2\%) &0 &$ -54.0764294 $&$0.00 $&$1.9502294[5]$& $-4.6013306[3]$& 7.283118  \\ 
&(96\%) &+ $1s^22p^4$ $^3P$ &(2\%) &1 &$ -54.0761997 $&$50.41 $&$1.9502254[5]$& $-4.6022305[3]$& 7.283118  \\ 
&(96\%) &+ $1s^22p^4$ $^3P$ &(2\%) &2 &$ -54.0758321 $&$131.07 $&$1.9502371[5]$& $-4.6055267[3]$& 7.283137  \\ 
$1s^22s^22p^2$ $^1D$ &(96\%) &+ $1s^22p^4$ $^1D$ &(2\%) &2 &$ -54.0059995 $&$15457.56 $&$1.9480134[5]$& $-4.4661335[3]$& 7.286404  \\ 
$1s^22s^22p^2$ $^1S$ &(92\%) &+ $1s^22p^4$ $^1S$ &(6\%) &0 &$ -53.9262127 $&$32968.75 $&$1.9454533[5]$& $-4.4612672[3]$& 7.279155  \\ 
$1s^22s2p^3$ $^5S^o$ &(99\%) &&&2 &$ -53.8639833 $&$46626.52 $&$1.9421309[5]$& $-8.9397181[3]$& 7.105985  \\ 
$1s^22s2p^3$ $^3D^o$ &(96\%) &+ $1s^22s^22p3d$ $^3D^o$ &(2\%) &3 &$ -53.6555256 $&$92377.69 $&$1.9351388[5]$& $-8.5991204[3]$& 7.105092  \\
\end{longtable}
\begin{table}[h]
\centering
\caption{\label{tab:Jon_C}Comparison of corrected ($K_{\rm SMS}$)  and  uncorrected ($K^1_{\rm SMS}$) specific shift parameters (in $m_e E_{\rm h}$) for several states of N~II, O~III, F~IV, Ne~V and Ti~XVII. See page \pageref{table_ISjon_C} for explanations.} % \medskip\\
\begin{tabular}{lccccccccccccccccc} \hline
&   & \multicolumn{3}{c}{N~II}   &&  \multicolumn{3}{c}{O~III} && \multicolumn{3}{c}{F~IV}\\ 
               \cline{3-5} \cline{7-9} \cline{11-13}
               &&\cite{JonBie:2010a}&\multicolumn{2}{c}{This work}&&
               \cite{JonBie:2010a}&\multicolumn{2}{c}{This work}&&
               \cite{JonBie:2010a}&\multicolumn{2}{c}{This work}\\
Level          & $J$ & 	$K_{\rm SMS}^1$ &$K_{\rm SMS}^1$ &$K_{\rm SMS}$&& 
				$K_{\rm SMS}^1$ &$K_{\rm SMS}^1$ &$K_{\rm SMS}$&& 
				$K_{\rm SMS}^1$ &$K_{\rm SMS}^1$ &$K_{\rm SMS}$ \\ 
\hline 
$2s^22p^2~^3P$ 	& 0 &$-$1.280&$-1.278$ & $-1.275$& & $-$2.473 & $-2.471$&$-2.461$& & $-$3.984 &$-3.983$&$-3.963$\\
               			& 1 &$-$1.279&$-1.277$ & $-1.275$& & $-$2.471 & $-2.469$&$-2.462$& & $-$3.981 &$-3.980$&$-3.964$\\
               			& 2 &$-$1.278&$-1.276$ & $-1.276$& & $-$2.468 & $-2.467$&$-2.464$& & $-$3.977 &$-3.976$&$-3.967$\\
$2s^22p^2~^1D$ 	& 2 &$-$1.236&$-1.239$ & $-1.237$& & $-$2.422 & $-2.423$&$-2.418$& & $-$3.926 &$-3.927$&$-3.915$\\
$2s^22p^2~^1S$ 	& 0 &$-$1.230&$-1.237$ & $-1.236$& & $-$2.415 & $-2.420$&$-2.414$& & $-$3.924 &$-3.927$&$-3.915$\\
\hline

\end{tabular}
\end{table}

\begin{table}[htdp]
\centering
\caption{\label{tab:C_MCHF_comp}Comparison of MCHF and MCDHF specific mass shift parameters for O~III (in atomic units $E_{\rm h} m_e$). }
\begin{tabular}{lcccccc@{}} 
\toprule
 &MCHF&&\multicolumn{3}{c}{MCDHF}\\
 \cline{4-6}
&\multicolumn{1}{c}{$K_{\rm SMS}^1$}&&$J$&$K^1_{\rm SMS}$&$K_{\rm SMS}$\\
\midrule
$1s^22s^22p^2$ $^3P$  	&  $-2.4877$		&&0&$-2.4710$&$-2.4614$\\
												&								&&1&$-2.4695$&$-2.4621$\\
												&  			  			&&2&$-2.4671$&$-2.4638$\\
$1s^22s^22p^2$ $^1D$	&	$-2.4316$		&&2&$-2.4232$&$-2.4181$\\
$1s^22s^22p^2$ $^1S$	&$-2.4171$	&&0&$-2.4195$&$-2.4144$\\
$1s^22s2p^3$ $^5S^o$   & $-4.3259$	&&2&$-4.3164$&$-4.3061$\\
$1s^22s2p^3$ $^3D^o$	&	$-4.2151$		&&3&$-4.2078$&$-4.1978$\\
					&			&&2&$-4.2077$&$-4.1974$\\
					&			&&1&$-4.2082$&$-4.1978$\\
$1s^22s2p^3$ $^3P^o$   	& $-4.0183$		&&2&$-4.0280$&$-4.0182$\\
					&			&&1&$-4.0275$&$-4.0174$\\
					& 			&&0&$-4.0276$&$-4.0175$\\
$1s^22s2p^3$ $^1D^o$	&	$-3.9245$ 		&&2&$-3.9200$&$-3.9094$\\
$1s^22s2p^3$ $^3S^o$   	& $-3.5915$	&&1&$-3.5903$&$-3.5786$\\
$1s^22s2p^3$ $^1P^o$	&	$-3.5399$		&&1&$-3.5825$&$-3.5720$\\
$1s^22p^4$ $^3P$		&	$-5.5479$		&&2&$-5.5704$&$-5.5532$\\
					&			&&1&$-5.5680$&$-5.5545$\\
					&			&&0&$-5.5666$&$-5.5549$\\
$1s^22p^4$ $^1D$		&	$-5.7031$		&&2&$-5.7273$&$-5.7124$\\
%mrg: I don't trust the next one $1s^22s^22p4p$ $^3P$	&	$-1.4236$		&&0&$-2.3391$&$-2.3288$\\
\bottomrule
\end{tabular}
\end{table}

\clearpage
\renewcommand{\arraystretch}{1.0}
\footnotesize
% !TEX root = adndt_IS_131121.tex

%\begin{table}[htdp]
%\caption{\label{tab:Thom_comp}Comparison of experimental and theoretical results ( ).}
%\begin{tabular}{@{}l|ccccccc} 
%&\multicolumn{3}{c}{N I}\\
%%&\multicolumn{2}{c}{$K_{\rm NMS}$}&
%&&\multicolumn{3}{c}{$K_{\rm SMS}$}\\
%%\cline{2-3}\cline{5-8}
%&Jennerich\cite{Jenetal:2006a}&Carette \cite{Caretal:2010b}&This work\\
%$3s~^4P_{1/2} \rightarrow 3p~^4D^o_{1/2}$&$-2488.1(15)$&$-2488.1(15)$\\
%$3s~^4P_{3/2} \rightarrow 3p~^4P^o_{1/2}$&$-2558.3(22)$&$-2579.4(68)$\\
%$3s~^4P_{5/2} \rightarrow 3p~^4D^o_{5/2}$&$-2748.17(84)$&$-2748.17(84)$\\
%$3s~^4P_{5/2} \rightarrow 3p~^4D^o_{7/2}$&$-2762.9(16)$&$-2746.4(18)$\\
%$3s~^4P_{5/2} \rightarrow 3p~^4P^o_{3/2}$&$-2713.4(14)$&$-2746.4(17)$\\
%$3s~^4P_{5/2} \rightarrow 3p~^4P^o_{5/2}$&$-2745.4(18)$&$-2745.4(18)$\\
%\end{tabular}
%%\end{indented}
%\end{table}

\begin{longtable}{@{\extracolsep\fill}llllllrccc@{}}
\caption{\label{tab:Nlike}Total energies (in $E_{\text h}$), excitation energies (in cm$^{-1}$), normal and specific mass shift $\widetilde{K}$ parameters (in GHz u), $\widetilde {\cal F}$ field shift factors (in GHz/fm$^2$) for levels in the nitrogen isoelectronic sequence ($7\leq Z\leq36$ and $Z=42,74$). For each of the many-electron wave functions, the leading components are given in the $LSJ$ coupling scheme. The number in square brackets is the power of 10. See page\ \pageref{table_ISN} for Explanation of Tables.}\\ [-.4cm]
%\multicolumn{4}{c}{Level composition in $LSJ$ coupling} & \multicolumn{1}{c}{$J$} & \multicolumn{1}{c}{Energy ($E_{\text h}$)} &\multicolumn{1}{c}{$\Delta E$ (cm$^{-1}$)}&  \multicolumn{1}{c}{$K_{\rm NMS}$ ($E_{\text h} m_e$)} &  \multicolumn{1}{c}{$K_{\rm SMS}$ ($E_{\text h} m_e$)}  &  \multicolumn{1}{c}{Density ($a_0^{-3}$)}   \\                                                                        
\multicolumn{4}{c}{Level composition in $LSJ$ coupling} & \multicolumn{1}{c}{$J$} & \multicolumn{1}{c}{Energy ($E_{\text h}$)} &\multicolumn{1}{c}{$\Delta E$ (cm$^{-1}$)}&  \multicolumn{1}{c}{$\widetilde K_{\rm NMS}$ (GHz u )} &  \multicolumn{1}{c}{$\widetilde K_{\rm SMS}$ (GHz u )}  &  \multicolumn{1}{c}{$\widetilde {\cal F}$~(GHz/fm$^{2})$} \\
\endfirsthead
\caption[]{(continued)} \\ [-.4cm]
\multicolumn{4}{c}{Level composition in $LSJ$ coupling} & \multicolumn{1}{c}{$J$} & \multicolumn{1}{c}{Energy ($E_{\text h}$)} &\multicolumn{1}{c}{$\Delta E$ (cm$^{-1}$)}&  \multicolumn{1}{c}{$\widetilde K_{\rm NMS}$ (GHz u )} &  \multicolumn{1}{c}{$\widetilde K_{\rm SMS}$ (GHz u )}  &  \multicolumn{1}{c}{$\widetilde {\cal F}$~(GHz/fm$^{2})$} \\
\hline
\endhead
\noalign{\vskip3pt}
\multicolumn{6}{r}{\itshape Continued\dots}\\
\endfoot
\noalign{\vskip3pt}\hline
\endlastfoot
\\
\hline&&&& \multicolumn{3}{c}{N~I} \\ \hline
$2s^22p^3$ $^4S^o$ &(96\%) &+ $2s2p^33d$ $^4S^o$ &(2\%) &3/2 &$ -54.6073794 $&$0.00 $&$1.9688775[5]$& $-6.3040379[3]$& 7.239658  \\ 
$2s^22p^3$ $^2D^o$ &(96\%) &+ $2s2p^33d$ $^2D^o$ &(1\%) &5/2 &$ -54.5185543 $&$19494.84 $&$1.9662690[5]$& $-6.1196713[3]$& 7.245065  \\ 
&(96\%) &+ $2s2p^33d$ $^2D^o$ &(1\%) &3/2 &$ -54.5185222 $&$19501.91 $&$1.9662598[5]$& $-6.1185845[3]$& 7.245077  \\ 
$2s^22p^3$ $^2P^o$ &(93\%) &+ $2p^5$ $^2P^o$ &(3\%) &3/2 &$ -54.4735477 $&$29372.66 $&$1.9648715[5]$& $-6.1771300[3]$& 7.241464  \\ 
&(93\%) &+ $2p^5$ $^2P^o$ &(3\%) &1/2 &$ -54.4735265 $&$29377.32 $&$1.9648663[5]$& $-6.1780280[3]$& 7.241484  \\ 
$2s2p^4$ $^4P$ &(91\%) &+ $2s^22p^23d$ $^4P$ &(3\%) &5/2 &$ -54.2058086 $&$88134.61 $&$1.9543893[5]$& $-9.7150959[3]$& 7.090616  \\
\end{longtable}

\begin{table}[htdp]
\centering
\caption{\label{tab:N_MCHF_comp}Comparison of MCHF and MCDHF specific mass shift parameters for  Ne~IV (in atomic units $E_{\rm h}m_e$).} 
\begin{tabular}{lcccrrr@{}} 
\toprule
 &MCHF&&\multicolumn{3}{c}{MCDHF}\\
 \cline{4-6}
&\multicolumn{1}{c}{$K_{\rm SMS}^1$}&&$J$&\multicolumn{1}{c}{$K^1_{\rm SMS}$}&\multicolumn{1}{c}{$K_{\rm SMS}$}\\
\midrule
$2s^22p^3$ $^4S^o$  	& $-8.1008$	&&3/2	&$-8.0918$&$-8.0570$\\
$2s^22p^3$ $^2D^o$	&	$-8.0133$	&&5/2		&$-8.0086$&$-7.9746$\\
					& 		&&3/2		&$-8.0083$&$-7.9732$\\
$2s^22p^3$ $^2P^o$	&	$-8.0209$	&&1/2		&$-8.0207$&$-7.9865$\\
					&		&&3/2		&$-8.0204$&$-7.9873$\\
$2s2p^4$ $^4P$    	  	& $-11.1298$	&&5/2		&$-11.1401$&$-11.0858$\\
					&		&&3/2		&$-11.1326$&$-11.0879$\\
                    &        &&1/2		&$-11.1292$&$-11.0901$\\
$2s2p^4$ $^2D$  		&$-10.9247$		&&5/2		&$-10.9608$&$-10.9123$\\
					&		&&3/2		&$-10.9599$	&$-10.9109$\\
$2s2p^4$ $^2S$		&$-10.5174$	&&1/2	&$-10.5672$&$-10.5186$\\
$2s2p^4$ $^2P$		&$-10.4440$		&&3/2		&$-10.4310$	&$-10.3761$\\
             		&        		&&1/2		&$-10.4228$&$-10.3800$\\
$2p^5$ $^2P^o$		&$-13.4948$		&&3/2		&$-13.5525$&$-13.4843$\\
					&		&&1/2		&$-13.5426$&$-13.4911$\\
\bottomrule
\end{tabular}
\end{table}
%\include{mass_shiftNlike}
%%
%%\normalsize\bigskip
%%References in the data tables may be either those given in the
%%introductory material or a separate list given following the
%%tables. An example of the formatting that could be used is shown above.
%%This list of references for the tables should begin on a new
%%page.

\end{document}